\documentclass{article}
\usepackage{amsmath, amsthm}
\def\br{\begin{eqnarray}}
\def\er{\end{eqnarray}}
\def\be{\begin{equation}}
\def\ee{\end{equation}}

\def\lb{\lbrack}
\def\rb{\rbrack}

\def\({\left(}
\def\){\right)}

\relax
\def\a{\alpha}

\def\b{\beta}

\def\d{\delta}

\def\g{\gamma}

\def\l{\lambda}
\def\L{\Lambda}

\def\pa{\partial}

\def\tp0{\Theta_{+}^{(0)}}
\def\tm0{\Theta_{-}^{(0)}}

\def\vp{\varphi}

                \def\a{\alpha}
                \def\b{\beta}

                \def\d{\delta}

                \def\g{\gamma}
                
                \def\vp{\varphi}

                \def\/{\frac}

                \def\l{\lambda}
                \def\L{\Lambda}

                \def\pa{\partial}

                \def\vp{\varphi}

                \def\({\Big(}
                \def\){\Big)}
                \def\[{\Big[}
                \def\]{\Big]}
  \def\rlx{\relax\leavevmode}
                \def\inbar{\vrule height1.5ex width.4pt depth0pt}
                \def\IZ{\rlx\hbox{\sf Z\kern-.4em Z}}
                \def\IR{\rlx\hbox{\rm I\kern-.18em R}}
                \def\IC{\rlx\hbox{\,$\inbar\kern-.3em{\rm C}$}}
                \def\IN{\rlx\hbox{\rm I\kern-.18em N}}
                \def\IO{\rlx\hbox{\,$\inbar\kern-.3em{\rm O}$}}
                \def\IP{\rlx\hbox{\rm I\kern-.18em P}}
                \def\IQ{\rlx\hbox{\,$\inbar\kern-.3em{\rm Q}$}}
                \def\IF{\rlx\hbox{\rm I\kern-.18em F}}
                \def\IG{\rlx\hbox{\,$\inbar\kern-.3em{\rm G}$}}
                \def\IH{\rlx\hbox{\rm I\kern-.18em H}}
                \def\II{\rlx\hbox{\rm I\kern-.18em I}}
                \def\IK{\rlx\hbox{\rm I\kern-.18em K}}
                \def\IL{\rlx\hbox{\rm I\kern-.18em L}}
                \def\one{\hbox{{1}\kern-.25em\hbox{l}}}
                \def\0#1{\relax\ifmmode\mathaccent"7017{#1}%
                B        \else\accent23#1\relax\fi}
                
\newtheoremstyle{theorem}
  {10pt}		  
  {10pt}  
  {\sl}  
  {\parindent}     
  {\bf}  
  {. }    
  { }    
  {}     
\theoremstyle{theorem}

\newtheoremstyle{defi}
  {10pt}		  
  {10pt}  
  {\rm}  
  {\parindent}     
  {\bf}  
  {. }    
  { }    
  {}     
\theoremstyle{defi}

\begin{document}
\pagestyle{empty}
\title{\bf Noncommmutative solitons and kinks in the affine Toda model coupled to matter}

\author{Harold Blas$^1$ and Hector L. Carrion$^2$\\
$^1$Instituto de F\'{\i}sica,
   Universidade Federal de Mato Grosso\\
   Av. Fernando Correa, s/n, Coxip\'o \\
   78060-900, Cuiab\'a - MT - Brazil\\
    blas@ufmt.br\\[2pt]
$^2$Instituto de F\'{\i}sica, Universidade de S\~ao Paulo,\\
   Caixa Postal 68528, 21941-972 S\~ao \,Paulo, Brazil.\\
   hlc@fma.if.usp.br}

\maketitle

\begin{abstract}
Some properties of the non-commutative (NC) versions of the
  generalized sine-Gordon model (NCGSG) and its dual massive Thirring
  theory are studied. Our method relies on the NC extension
  of integrable models and the master lagrangian approach to deal with
  dual theories. The master lagrangian turns out to be the NC version
  of the so-called affine Toda model coupled to matter
  related to the group $GL(n)$, in which the Toda field $g \subset GL(n),\,(n=2, 3)$. Moreover, as a reduction of $GL(3)$ NCGSG one gets a NC version of the remarkable double sine-Gordon model.


\end{abstract}
\pagestyle{empty}
\section{{\bf Introduction}}

Some non-commutative versions of the sine-Gordon model (NCSG) have
been proposed in the literature
\cite{lechtenfeld, jhep2, grisaru1}. The relevant equations of
motion have the general property of reproducing the ordinary
sine-Gordon equation when the non-commutativity parameter is
removed. It has been shown that these models arise as reduced models starting from a master Lagrangian, the so-called $sl_{2}$ affine Toda model coupled to matter \cite{jhep2}. In the present paper we summarize those results, as well as certain new NC versions of the (generalized) sine-Gordon and massive Thirring dual models  \cite{jhep1}.

In the following we consider an algebra of continuous functions with the Moyal product or star product
\br {f} \star {g } (x) &=& \mbox{exp}\(\frac{i}{2} \theta^{\mu \nu}
\partial_{\mu}^{(x')} \partial_{\nu}^{(x'')} \) {f}(x')
{g}(x'') |_{x'=x''=x}. \er

So, one has the coordinate noncommutativity, i.e.,  $\[\hat{x}^{\mu}, \hat{x}^{\nu} \]_{\star}= i
\theta^{\mu \nu}$.

\section{{\bf Sine-Gordon/massive Thirring models}}

The sine-Gordon action is defined by
\br
 S = \frac{1}{2} \int \,\, d t\,d x \bigl[
\pa_\mu\vp\pa^{\mu}\vp + \frac{2\a_{0}}{\b^2} (\cos \b \vp-1) \bigr]
\er
for a scalar field~$\vp(t,x)$ on $\IR^{1,1}$.
This model has many remarkable features, such as
a Lax-pair representation,
infinitely many conserved local charges,
a factorizable S-matrix,
as well as soliton and breather solutions.
The simplest soliton configuration is
\br \label{kink}
 \vp_{\mbox{kink}}(t,x) = 4 \mbox{arctan}\,\mbox{exp}[-\frac{\a_{0}}{2} \xi]\er
$\mbox{with}\,\,  \xi =\frac{x-v t}{\sqrt{1-v^2}}$  .

In the usual space-time the quantum equivalence between the massive Thirring and sine-Gordon models is well known. The Thirring model
\br S_{NCMT} &=& \int d t d x \bigl[ i\bar{\psi} \g^{\mu} \pa_{\mu}  \psi
+ m \bar{\psi}  \psi -\frac{\l}{2} j^{\mu}
j_{\mu} \bigr]\nonumber\\
&&  j^{\mu} = \bar{\psi} \g^{\mu} \psi \label{mt1}\er
is equivalent to the sine-Gordon model provided that $
\frac{4\pi}{\b^2} = 1 +  \frac{\l}{\pi}$. The same equivalence holds in NC space-time even though their actions look very different. However, the coupling  relationship maintains its form up to some re-scalings \cite{duality}. In the following we provide the classical NC counterparts of the SG and MT models, as well as some multi-field generalizations.

\section{{\bf The master lagrangian}}

The NC affine Toda model coupled to matter (spinor) (NCATM) is defined by
\br
S_{NCATM} & \equiv &  S[g, W^{\pm},F^{\pm}]
 = I_{WZW}[g] + \nonumber \\ &&  \int d^2x \{ \frac{1}{2} Tr [
\partial_{-} W^{-}\star [E_{2}\,,\, W^{-} ]] - \nonumber \\ && \frac{1}{2}
Tr[ [E_{-2}\,,\,W^{+}] \star \partial_{+} W^{+} ] + \nonumber \\
&& Tr[ F^{-} \star \partial_{+} W^{+} ] + Tr[ \partial_{-} W^{-}
\star F^{+}] +  \nonumber \\ && Tr [ F^{-}\star g \star F^{+}
\star g^{-1} ] \}, \label{sncatm1}
\er
where
$F\star G =
F\,\mbox{exp}\(\frac{\theta}{2}(\overleftarrow{\pa_{+}
}\overrightarrow{\pa_{-}}-\overleftarrow{\pa_{-}}
\overrightarrow{\pa_{+}})\) G$,
{\bf $E_{\pm2}$} is a constant matrix. Consider $ x_{\pm}=t\pm x,\,$ then $,\,
\pa_{\pm}=\frac{1}{2}(\pa_{t}\pm \pa_{x})$.
$I_{WZW}[g]$ is the {\bf NC Wess-Zumino-Witten} model
\br
I_{WZW}[g] &=& \int d^2x
\pa_{+} g \star \pa_{-} g^{-1} + \nonumber \\ && \int_{0}^{1} dy
\hat{g}^{-1} \star \pa_{y} \hat{g} \star
\[\hat{g}^{-1} \star \pa_{+}  \hat{g}, \hat{g}^{-1} \star \pa_{-}
\hat{g} \]_{\star} , \nonumber
\er
 where $\hat{g}(y)$ is such that $\hat{g}(0)=1$,
$\hat{g}(1) = g$ ($[y,x_{+}]=[y,x_{-}]=0$)

The matrix fields are defined by
\br \nonumber E_{\pm 2}= \frac{m_{\psi}}{4} H^{\pm
1},\,\, \,
g[SG\,\,\mbox{field(s)}] \in  U(1) \times U(1)\,\,\, \mbox{or} \,\,\,  U(1)_{\IC}\nonumber\er
\br  F^{+}=
\sqrt{i m_{\psi}} \( \psi_{R}E_{+}^{0}+ \widetilde{\psi}_{R}
E_{-}^{1} \),\,\,F^{-} = \sqrt{im_{\psi}} \(\psi_{L}
E_{+}^{-1}-\widetilde{\psi}_{L} E_{-}^{0}\). \er
The affine Lie algebra $\hat{sl}(2)$ commutation relations are
\br
 \lb H^m \, , \, H^n \rb &=& 2 \, m \, C \, \d_{m+n,0},  \label{sl2a}\\
                \lb H^m \, , \, E^n_{\pm} \rb &=& \pm 2 \, E^{m+n}_{\pm},
                \label{sl2b}\\
   \lb E^m_{+} \, , \, E^n_{-} \rb &=& H^{m+n} + m \, C \, \d_{m+n,0},
                \label{sl2c}
  \er
The Dirac fields in components become
  \br
  \psi = \(\begin{array}{l}
\psi_{R} \\
\psi_{L}
\end{array} \);\,\,\,\,\widetilde{\psi} = \(\begin{array}{l}
\widetilde{\psi}_{R} \\
\widetilde{\psi}_{L}
\end{array} \)
  \er

The equations of motion for the NCATM model in matrix form are
  \br  \pa_{+}(
\pa_{-} g \star g^{-1}) &=& \[F^{-}\,,\, g\star F^{+}\star g^{-1}\] \nonumber
\\
\partial_{+} F^{-} = - [E_{-2}, g \star F^{+} \star g^{-1}]&,&
\partial_{-} F^{+} =-[E_{2}, g^{-1} \star F^{-} \star g ].\nonumber
  \er

A `local' symmetry of the model \cite{hb1, jhep1} allows one to view the field
$ g $ as a `gauge field' and
$F^{\pm}$ as matter
fields. Due to that symmetry the NCATM model can be reduced, through some methods, to the NCSG$_{1, 2}$ models.\\

\subsection{{\bf NC versions of the sine-Gordon model $NCSG_{1,2}$}}

The general effective action of the reduced bosonic model becomes
 \begin{eqnarray}
  S[g] =  I_{WZW} [g] +\int d^2x \{ Tr[ \Lambda^{-}  \star g \star \Lambda^{+}
                \star g^{-1}]\}. \label{ncsg1}
                \end{eqnarray}
 {\bf First Version.} Consider $g \in U(1)$x$U(1)$ in the representation \br
\label{u1u1} g = \left(\begin{array}{cr}
e^{i\vp_{+}}_{\star} &  0 \\
0 & e^{-i\vp_{-}}_{\star}
\end{array} \right)\,\,,
g_{+} \equiv \left(\begin{array}{cr}
e^{i\vp_{+}}_{\star} &  0 \\
0 & 1
\end{array} \right),\,\,\, g_{-} \equiv \left(\begin{array}{cr}
1 &  0 \\
0 & e^{-i\vp_{-}}_{\star}
\end{array} \right)\nonumber
\er with $\vp_{\pm}$ being real fields.

For the $\Lambda$'s taken as
   \br \nonumber
   \Lambda^{+}=
   M(E_{+}^{0}+E_{-}^{1}),\,\,\,\Lambda^{-}= M(E_{-}^{0}+E_{+}^{-1}),
   \er
the action (\ref{ncsg1}) can be written as
\br\label{nclech11}
                   S_{NCSG_{1}}[g_{+}, g_{-}] && = I_{WZW}[g_{+}] +
                   I_{WZW}[g_{-}] +  \nonumber \\ && M^{2} \int d^2x
\( e_{*}^{i \vp_{+}+ i \vp_{-}}+ e_{*}^{-i \vp_{+}- i \vp_{-}} -2\).
   \er
The eqs. of motion become \br\label{ncsgeq1} \nonumber
\pa_{+}\( \pa_{-} e_{\star}^{i
\pm \vp_{\pm}} \star  e_{\star}^{\mp i\vp_{\pm}}\)= \mp M^{2} \(e^{i\vp_{-}}_{\star} \star e_{\star}^{i
\vp_{+}}- e^{-i\vp_{+}}_{\star} \star e_{\star}^{-i \vp_{-}}\)\label{ncsgeq2}. \nonumber
\er
This is the Lechtenfeld et. al. proposal for NCSG$_{1}$ model. In the $\theta
\rightarrow 0$ limit one has  a free field eq., $ \label{limit1}\pa_{-}\pa_{+} \(\vp\)=0$;\, $(\vp\equiv \vp_{+}-\vp_{-})$\,\, and the usual SG  eq.
 $\pa^{2} \vp_{SG} = -4 M^2 \mbox{sin} (\vp_{SG})$ ($\vp_{SG} \equiv \vp_{+}+\vp_{-}$).\\

{\bf Second Version.} Consider $g \in U(1)_{\IC}$ \br \label{exp}
                           g= e^{i\vp H^{0}}_{\star} \equiv
                           \left(\begin{array}{cr}
                           e^{i\vp}_{\star} &  0 \\
                           0 & e^{-i\vp}_{\star}
                           \end{array} \right)\,\,\,\,\, \mbox{and}\,\,\,\,\,
                           \bar{g}= g^{\star}, \er
where the field $\vp$ is a general complex field. The equations of motion are
\br \pa_{+}\( \pa_{-} e^{\mp \frac{i}{2} \vp} \star e^{\pm \frac{i}{2} \vp} \)= \pm \frac{\g}{4}\(e^{i \vp} - e^{-i\vp}\),\,\,\g=const.
\er
and similar equations for $\bar{g}$ considered as an independent field. This is the Grisaru-Penati proposal for the NCSG$_{2}$.\\

\section{\bf{NC versions of the massive Thirring model}}

The dual to the NCSG$_{1}$ model turns out to be the NC bi-fundamental $U(1)\times U(1)$ massive Thirring
model \cite{jhep2, duality}.  The (Euclidean) Lagrangian is
\br {\cal L}_{NCMT_{1}}=  i\bar{\psi} \g^{\mu} \pa_{\mu} \star \psi
+m \bar{\psi} \star \psi -\frac{\l_{b}}{4} j^{(A)\,\mu} \star
j^{(A)}_{\mu} -  \frac{\l_{b}}{4} j^{(B)\,\mu}\star j^{(B)}_{\mu}
 \label{ncmt0}, \er
defined for Dirac fields and with the currents given by
\br j^{(A)\,\mu} \equiv \bar{\psi}\star \g^{\mu}
\psi;\,\,\, j^{(B)\,\mu} \equiv \psi_{\b} (
\g^{\mu})_{\a\b} \star \bar{\psi}_{\a}.\er
 Here, $\l_{b}$ is the
coupling constant and the group index contractions are being
assumed. The currents correspond to the $U(1)\times U(1)$ symmetry in NC space. Two copies of the above NCMT$_{1}$ model defines the relevant NCMT$_{2}$ corresponding to the NCSG$_{2}$.\\
\section{{\bf The generalized NC sine-Gordon NCGSG$_{1}$}}

 Let us consider $g \subset GL(n),\, n>2$. We must take the bosonic action (\ref{ncsg1}) for a relevant set $\{\L^{\pm}_{m},\, g\}$\cite{hb1}. Consider the parametrization \,$g \in [U(1)]^3 \subset GL(3,
\IC)$
\br
\label{u1u1u1} g = \left(\begin{array}{ccc}
e^{i\phi_{1}}_{\star} & 0&  0 \\
0 & e^{i\phi_{2}}_{\star} & 0 \\
0 & 0 & e^{i\phi_{3}}_{\star}  \\
\end{array} \right)\,\equiv\, g_{1}* g_{2} * g_{3},
\er
with $\phi_{i}$ being real fields ($i=1,2,3$). The equations of motion are
\br\nonumber
\pa_{+}\( \pa_{-} e_{\star}^{i
\phi_{i}} \star e_{\star}^{-i\phi_{i}}\) =  G_{i},\,\,\,\,i=1,2,3\,\,\,\,\,\,\,\,\,\,\,\,\,\,\,\,\,\,\,\,\,\,\,\,\,\,\,\,\,\,\,\,\,\\
G_{1} \equiv \frac{M_{1}}{8}[
e_{\star}^{i \phi_{2}} \star e_{\star}^{-i \phi_{1}}-
  e_{\star}^{i \phi_{1}} \star e_{\star}^{-i\phi_{2}}] +\frac{M_{3}}{8}[ e_{\star}^{i
\phi_{3}} \star e_{\star}^{-i \phi_{1}}-
 e_{\star}^{i\phi_{1}} \star e_{\star}^{-i\phi_{3}}]\nonumber\\
G_{2} \equiv  \frac{M_{2}}{8} [
e_{\star}^{i \phi_{3}} \star e_{\star}^{-i \phi_{2}}-
 e_{\star}^{i \phi_{2}} \star e_{\star}^{-i \phi_{3}}] + \frac{M_{1}}{8}[ e_{\star}^{i
\phi_{1}} \star e_{\star}^{-i \phi_{2}}- e_{\star}^{i \phi_{2}}
\star e_{\star}^{-i \phi_{1}}].\nonumber\\
G_{3} \equiv  \frac{M_{3}}{8} [
e_{\star}^{i \phi_{1}} \star e_{\star}^{-i
\phi_{3}}- e_{\star}^{i
\phi_{3}} \star e_{\star}^{-i \phi_{1}}] + \frac{M_{2}}{8}[  e_{\star}^{i
\phi_{2}} \star e_{\star}^{-i \phi_{3}}- e_{\star}^{i \phi_{3}} \star e_{\star}^{-i
\phi_{2}}].\nonumber \er
This  system defines the  NCGSG$_{1}$.
In the $\theta \rightarrow 0$ limit one gets a free scalar \br \pa^2\,\Phi&=&0, \,\,\,\,\,\,\Phi \equiv
\phi_{1}+\phi_{2}+\phi_{3}. \nonumber\er

For the particular solution $\Phi\equiv 0$, one can write
\br
\nonumber\pa^2\,\phi_{1}&=&  M_{1}
\sin(\phi_{2}+\phi_{1})+
M_{3}\sin(2\phi_{1}-\phi_{2});\\
\pa^2\,\phi_{2}&=&  M_{2} \sin(2\phi_{2}-\phi_{1})+ M_{1}
\sin(\phi_{1}+\phi_{2} ). \er

This is the usual generalized
sine-Gordon model \cite{hb3}. Remarkably, a version of the  NC double sine-Gordon model (NCDSG$_{1}$) emerges for the reduction
 $ \phi_{1} = -\phi_{3} =
\phi,\;\phi_{2}=0$. So, one gets the equations \br
\pa_{+}\( \pa_{-} e^{i \phi}_{\star} \star e^{-i \phi}_{\star}
+ \pa_{-} e^{-i \phi}_{\star}
\star e^{i \phi}_{\star} \)&=&0,\label{zero}\\
 \pa_{+}\( \pa_{-} e^{-i
\phi}_{\star} \star e^{i \phi}_{\star}\) &=& 2i M_{1}\, \mbox{sin}_{\star} \phi + 2i
M_{3}\,
\mbox{sin}_{\star} 2 \phi; \label{ncdsg}\er

In the limit $\theta \rightarrow 0$ eq. (\ref{ncdsg}) reduces to
$
\pa_{+} \pa_{-}  \phi = - 2 M_{1}\, \mbox{sin} \phi - 2
M_{3}\,
\mbox{sin} 2 \phi $.

A second version NCGSG$_{2}$ exists for an another form of  $g$, see \cite{hb1}.\\

\section{{\bf Generalized NC Thirring model NCGMT$_{1}$}}

The dual to the NCGSG$_{1}$ model defined for three Dirac fields becomes \cite{hb1}
\br\nonumber {\cal L}_{NCGMT_{1}}&=&  i\sum_{j=1}^{3} \(\bar{\psi}_{j} \g^{\mu} \pa_{\mu} \star \psi_{j}
+ m_{j} \bar{\psi}_{j} \star \psi_{j}\) - \\ && \sum_{j}^{3} g_{jj} \(j^{(A)\,\mu}_{j} \star
j^{(A)}_{\mu\, j} + j^{(B)\,\mu}_{j} \star j^{(B)}_{\mu\, j}\)\nonumber\\
&&+ g_{12} \Big[j^{(A)\,\mu}_{1} \star
j^{(B)}_{\mu\, 2}\Big] - g_{23}\Big[j^{(A)\,\mu}_{2}\star
j^{(A)}_{\mu\, 3}\Big] - g_{13}\Big[ j^{(B)\,\mu}_{1} \star
j^{(B)}_{\mu\, 3}\Big]. \nonumber \er
\br j^{(A)\,\mu}_{j} \equiv \bar{\psi}_{j}\star \g^{\mu}
\psi_{j};\,\,\, j^{(B)\,\mu}_{j} \equiv \psi_{j\, \b} (
\g^{\mu})_{\a\b} \star \bar{\psi}_{j\, \a}. \nonumber \er
 Here, $g_{jk}$ are the
coupling constants and the group index contractions are being
assumed.

A second version NCGMT$_{2}$ , corresponding to the NCGSG$_{2}$ mentioned above, can be obtained by doubling the number of Dirac fields.\\

\begin{center}
\section{{\bf Discussion}}
\end{center}

The solitons and kinks of the NC sine-Gordon/massive Thirring models have been constructed in \cite{hb3}. It is known that if $f(t,x)$ and $g(t, x)$ depend only on $(x- v t)$, then $f \star g \Rightarrow f g $. Therefore, all the $\star$ products for this type of functions (e.g.  soliton and kink type solutions) become the same as the ordinary ones. So, $1-$soliton solutions of the usual GSG model are also solutions of the NC models. The NC extension of the SG/MT duality provides a rich structure already at the classical level. The NC actions differ from their commutative counterparts. By suitably reducing the NCGSG$_{1, 2}$ models one can get some versions of the NC double sine-Gordon model (NCDSG$_{1, 2}$). Moreover, the $1-$kink solution of the DSG model solves its NC counterparts NCDSG$_{1, 2}$. Several future research directions arise: multi-solitons, soliton scattering, quantum versions, S-matrix, and hopefully some physical applications.

\end{document}